\documentclass[runningheads]{llncs}
\usepackage[T1]{fontenc}
\usepackage{graphicx}
\usepackage{url}

\usepackage{breakurl}
\usepackage[breaklinks]{hyperref}
\usepackage{graphicx}
\begin{document}
\title{Turning to Online Forums for Legal Information: A Case Study of GDPR’s Legitimate Interests}
\titlerunning{Turning to Online Forums for Legal Information}
%
\author{Lin Kyi\inst{1}\orcidID{0000-0002-4753-3889} \and
Cristiana Santos\inst{2}\orcidID{0000-0003-0712-2038} \and
Sushil Ammanaghatta Shivakumar\inst{1}\orcidID{0000-0002-7042-7232} \and
Franziska Roesner\inst{3}\orcidID{0000-0001-8735-4810} \and
Asia Biega\inst{1}\orcidID{0000-0001-8083-0976}} 
\authorrunning{Kyi et al.}
%
\institute{Max Planck Institute for Security and Privacy, Bochum, Germany \and
Utrecht University, Utrecht, Netherlands \and
University of Washington, Seattle, United States}
%
\maketitle              
\begin{abstract}
Practitioners building online services and tools often turn to online forums such as Reddit, Law Stack Exchange, and Stack Overflow for legal guidance to ensure compliance with the GDPR. The legal information presented in these forums directly impacts present-day industry practitioner’s decisions. Online forums can serve as gateways that, depending on the accuracy and quality of the answers provided, may either support or undermine the protection of privacy and data protection fundamental rights. 
However, there is a need for deeper investigation into practitioners' decision-making processes and their understanding of legal compliance when seeking for legal information online. 

Using GDPR's ``legitimate interests'' legal ground for processing personal data as a case study, we investigate how practitioners use online forums to identify common areas of confusion in applying legitimate interests in practice, and evaluate how legally sound online forum responses are.

Our analysis found that applying the legal basis of legitimate interest is complex for practitioners, with important implications for how the GDPR is implemented in practice. The legal analysis showed that crowdsourced legal information tends to be legally sound, though sometimes incomplete.  
We outline recommendations to improve the quality of online forums by ensuring that responses are more legally sound and comprehensive, enabling practitioners to apply legitimate interests effectively in practice and uphold the GDPR.

\keywords{GDPR \and Legitimate Interest \and Online Forums \and Legal Information 
 \and Compliance}
\end{abstract}
\section{Introduction}
\label{sec:intro}
The General Data Protection Regulation (GDPR) came into effect in May 2018 to protect EU users' personal data~\cite{EUdataregulations2018}. 
However, it also brought  challenges for organizations who must translate abstract legal principles into specific technical requirements to assure compliance~\cite{bygrave2017data,finck2021reviving,horstmann2024those,jasmontaite2018data,shanmugam2022learning}.

How then, do practitioners (anyone involved in applying the GDPR for their respective organization, including  developers, designers, business owners, and others), particularly those without in-house legal support, navigate  GDPR obligations? While some official compliance guidance by regulatory authorities exists, some practitioners, also referred to as posters in this paper\footnote{Here, \textit{posters} refer to the practitioners who are posting questions in online forums. We also use the term \textit{practitioners} when referring to the same users posting questions too.}, may turn to online forums to seek legal information.
We use the term ``legal information''\footnote{There is a difference between ``legal information'' and ``legal advice.'' When posters refer to their concrete cases on Stack Overflow, they are provided with legal information tuned to their context. Legal advice is instead given by a legal professional and establishes an attorney-client relationship, which is not the case in online forums. In this paper, we use the term ``information'' to signify the legal case-related recommendations about a given post.} to signify the legal case-related recommendations in a given post. Online forums can provide scenario-specific, cost-free, and quick access to answers, and their use by industry practitioners has been documented in other contexts~\cite{fischer2017stack,tahaei2022understanding,wu2019developers}. 
Online forums can serve as gateways that, depending on the accuracy and quality of the answers provided, may either support or undermine the protection of privacy and data rights through adherence to data protection regulations.

In this paper, we investigate online forums in the context of GDPR compliance. We look at how practitioners seek legal information using the following online forums -- Reddit, Law Stack Exchange, and Stack Overflow -- to be compliant with the GDPR.

We explore online forums for the following reasons.
First, developers tend to primarily use online forums to support their privacy-related decisions~\cite{parsons2023understanding,tahaei2020understanding}. Developers often struggle to meet legal requirements as they often lack privacy and legal expertise~\cite{horstmann2024those,horstmann2025sorry}. 
Second, other tools to support their decision-making processes, such as official and reliable guidance for developers, are largely lacking~\cite{CNIL_GDPR_Developers_Guide,EDPB_SME_Data_Protection_Guide}. These guides are not contextually-sensitive for developer's projects, thus they are left with industry-based sources~\cite{Taylor-Hiscock_2021_GDPR_Compliance} which, while might be practical, are typically unofficial, potentially biased, and not subject to rigorous legal or regulatory review.
%
Consequently, the legal information presented in these forums directly impacts present-day industry practitioner’s decisions, given the prevalence of forums to aid developers~\cite{tahaei2020understanding}. 
With the development of Large-Language Models (LLMs), a novel source to access online legal information has been introduced into the software development process. 
This shift is reflected in trends, such as Stack Overflow reporting a decline in user activity, which they attributed to the rise of AI systems~\cite{da2025llms}. 
Third, more practitioners and users have started using LLMs and other AI assistants for legal information seeking and use~\cite{law_gpt,da2025llms} which are trained on online forum data~\cite{hamalainen2023evaluating}. 
Incorrect information present in these forums will thus have even bigger consequences with the usage of these new technologies~\cite{aircanada}. Therefore, it is important that online forums provide correct legal information for current and potential future uses of legal information.

The GDPR offers an extensive data protection regime that yields thousands of posts and responses when searched for in online forums. To make data analysis more feasible, we decided to focus our investigation specifically on the questions and responses that practitioners posed relating to “legitimate interests,” one of the GDPR’s six legal grounds for collecting data. We focused on legitimate interest because of its vague nature, yet is highly important due to its flexibility and breadth, covering legal bases for data collection not currently mentioned by the GDPR~\cite{ferretti2014data}.

Legitimate interest is defined in Art. 6(1)(f) of the GDPR as the processing that is necessary for the legitimate interests pursued by data controllers~\cite{santos2021consent} or third parties~\cite{EUdataregulations2018,EDPB-6-14}. While consent is generally well-established and recognized due to the proliferation of cookie banners, legitimate interest is not~\cite{kyi2023investigating}. In this paper, we analyze online forum discussions relating to legitimate interests to understand what confusions practitioners may have in applying it, and assumptions practitioners have about this legal basis.

The vague nature of legitimate interests has led to documented examples of its misuse and non-compliance~\cite{ferretti2014data,kyi2023investigating}. We hypothesize that one potential reason for legitimate interest misuse lies in the difficulty of understanding it.

\textbf{Contributions.} Our paper presents two main contributions. First, through a qualitative analysis of the questions in the online forums, we identified the main challenges practitioners may face when applying the legitimate interest legal basis ``in the wild'' by studying online forum posts. Second, our analysis of forum answers offering legal information indicates that while information may seem legally sound, there are precautions practitioners must take before relying on these forums. In our paper, we offer recommendations for how online legal forum posts could be formulated to ensure legally sound answers, therefore helping to better protect users’ fundamental rights to privacy and data protection.

\section{Background}
\label{sec:background}

\subsection{Empirical studies on the legitimate interest legal basis}
Court cases have shown that many uses of legitimate interests are illegal when used for targeted advertising purposes~\cite{irishdpa}, supplemented by empirical studies that have investigated the use of this legal basis. Matte et al.~\cite{matte2020purposes} found that hundreds of advertisers relied on legitimate interests for purposes that should instead rely on consent~\cite{matte2020purposes}. 
Kyi et al. found that in the context of cookie banners, the use of legitimate interests are not transparent to users: very few websites mentioned that they relied on legitimate interests, and of those which did, not all allowed users to object thereto. Their user study revealed that users are not fond of most legitimate interest-based purposes, such as personalized advertising, and preferred sharing data for purposes which did not benefit companies and advertisers~\cite{kyi2023investigating}. 


\subsection{Importance of Online Forums for Software Development}


\label{sec:onlineforumsfordevs}
Online forums are an important source of information for software developers~\cite{bacchelli2012harnessing,barua2014developers,parnin2012crowd}. Previous research has shown that Stack Overflow influences developers' practices, such as code reuse from Stack Overflow posts~\cite{wu2019developers}, and impacts developers' productivity on GitHub~\cite{vasilescu2013stackoverflow}. However, they may also promote poor practices such as reusing code, which has negative impacts on the security and privacy of systems~\cite{fischer2017stack}. 

The number of GDPR-related Stack Overflow posts in recent years has been increasing~\cite{tahaei2020understanding}, which suggests that privacy and data protection are becoming more important for developers~\cite{tahaei2022privacy}. The knowledge shared in online forums can impact how developers abide to security and data protection obligations~\cite{acar2016you,fischer2017stack,li2021developers,tahaei2020understanding}. Privacy-related discussions on Stack Overflow seem to be connected to external events relevant to privacy, such as new privacy restrictions~\cite{tahaei2022privacy,li2021developers}. When such an event happens, developers often felt privacy restrictions required more efforts from them without many benefits~\cite{li2021developers}.

\subsubsection{Online legal compliance information}
\label{sec:onlinelegalinformation}

The most common privacy-related information that developers suggested on Stack Overflow refers to legal compliance, as 38.7\% of answers related to GDPR and California Consumer Privacy Act (CCPA) compliance according to a study conducted by Tahaei et al.~\cite{tahaei2022understanding}. 
Developers found it difficult to translate legal requirements, which some saw as full of ``legalese,'' into technical terms~\cite{bednar2019engineering,greene2018platform,senarath2018developers,spiekermann2012challenges,tahaei2021privacy}. Developers frequently asked about how to adhere to legal requirements that are imposed by various platforms, such as different app stores. It was commonly advised for developers to check that their company's privacy policy was compliant, and to ask for user consent~\cite{tahaei2022understanding}.

Several online resources have emerged, aimed at helping industry practitioners, often developers, apply the GDPR. Some are from Data Protection Authorities (DPAs), such as the French DPA's ``GDPR Developer's Guide''~\cite{CNIL_GDPR_Developers_Guide} and European Data Protection Board's ``Data Protection Guide for Small Business''~\cite{EDPB_SME_Data_Protection_Guide}. Other sources of online information derive from developers aimed at helping fellow developers~\cite{GDPR4Devs}, and GDPR compliance guides from Consent Management Platforms (CMPs), such as OneTrust~\cite{Taylor-Hiscock_2021_GDPR_Compliance}. 

\section{Methods}
\label{sec:methods}
As the topic of our analysis spans questions at the intersection of technology and law, we collected data from three popular online forums that technology practitioners may turn to for advice: Reddit, Law Stack Exchange, and Stack Overflow. Our general approach was to collect and analyze this data to understand the struggles and discussions practitioners are having online when it comes to GDPR compliance, and to evaluate the legal soundness of the answers. 


We split our qualitative analysis into two sections to investigate i) the possible elements of confusion about GDPR compliance, specifically focusing on legitimate interests (Section \ref{sec:results:practitioners}), and ii) the legal soundness of online forum answers (Section \ref{sec:results:legalsoundness}). 

%
%
%
\subsection{Data Collection}
\paragraph{Reddit.}
\label{sec:reddit}
We searched for posts and comments from industry practitioners containing the phrase “legitimate interest(s)” on September 27, 2024. We made requests to the Reddit API to collect relevant data for our analysis, filtered results based on the subreddit, and stored this data in a CSV file. Our scrape collected the title of the post, URL (in case further analysis was needed), comments, the post’s body text, subreddit name, upvotes for the post, age of the post, and word count. We manually removed some posts from our dataset which were out of the scope of this study (e.g., describing legitimate interests in offline contexts). Most of our posts came from the r/GDPR and r/privacy subreddit due to the relevance of these posts for our analysis.
%
%

\paragraph{Law Stack Exchange and Stack Overflow.}
\label{sec:stackexchange}
Similar to Reddit, we searched for posts and comments containing “legitimate interest(s)” on Stack Overflow and Law Stack Exchange and used data from Stack Exchange’s public data dump from May 2024. We paired the posts with their comments and answers, and cleaned the data. We collected the title of the post, body text, URL, number of views, upvotes, comments, age of the post, and word count of the post.

\subsection{Dataset}
\label{sec:dataset}
In total, we collected 319 posts (not including comments and answers); 10 posts were from Stack Overflow, 203 posts from Law Stack Exchange, and 106 from Reddit, a sample size on par with previous qualitative papers analyzing online forums in privacy-specific contexts~\cite{tahaei2022understanding,tahaei2022privacy} and those used in qualitative legal research~\cite{Webley-qualit-content-analysis,Liepia2019Claudette}.
For the legal soundness analysis, we collected the answers to questions from Law Stack Exchange which were marked by posters as ``Accepted,'' which yielded 94 answers. We discuss the data analysis methods for our two-part analysis in each respective section. 

\noindent \textbf{Ethical Considerations of Using Online Forum Data.}
\label{sec:ethics}
Social media data can be rich and allow researchers to understand social phenomena, but also presents several ethical challenges~\cite{beadle2025sok}. A study using social media data was preferred over other research methods, such as interviewing or surveying industry practitioners, because: i) it would be difficult to gather industry participants who would be willing to discuss their company's data practices in a research context, ii) this data is readily available for analysis, and iii) online forums allow researchers to view how practitioners might navigate legal compliance ``in the wild''. 
A social media analysis would ensure a more targeted and scalable study of what industry practitioners are discussing for technical implementations of legitimate interest.
Reddit and Stack Exchange (including Stack Overflow) are two commonly-used sources of data for academic research~\cite{proferes2021studying,stackexch}. 
Due to the potential legal and ethical implications of these posts (such as an employee or company getting into legal troubles), when quoting participants, we did not mention usernames, and took precautions to paraphrase the posts. Sometimes we engaged in \textit{ethical fabrication} to maintain user anonymity when posters mentioned specific scenarios that could possibly be identifiable~\cite{barakat2022community,dym2020ethical,markham2012fabrication}. Although Stack Exchange prefers that researchers attribute posts~\cite{attribution}, due to the potential legal and ethical implications, we paraphrased quotes and did not link them to protect user identities.


\section{Analysis 1: Practitioner Discussions and Inquiries}
\label{sec:results:practitioners}
In this section we report on the qualitative analysis of the \textit{posts} we collected from the three forums. We investigated the practitioner roles seeking legal information, and common points of discussion (or confusion) about applying legitimate interests in practice. 


\noindent \textbf{Qualitative Analysis.} We analyzed the content from the three forums and conducted a \textit{thematic analysis} to find underlying patterns within the dataset by labeling important parts in the data (\textit{codes}), and grouping these codes to form \textit{themes}~\cite{braun2006using}. To investigate what practitioners discuss in online forums, we had two annotators analyse the collected posts. 
They first annotated a random set of 15 posts together to discuss what to focus the analysis upon. They decided to keep track of: 
i) the (perceived) role or industry of the poster, ii) the general topic they were seeking information about, and iii) codes related to the poster's query. 
After this was established, the annotators coded the same random set of 15 posts to calculate the interrater reliability. The agreement rate was 85\%, with a Cohen's $\kappa =$ 0.66, which indicates substantial agreement~\cite{mchugh2012interrater}. Due to the high agreement, the annotators split the dataset, and annotated the remaining half (107 posts) on their own.
In total, 135 codes were identified from the dataset, which fit into nine major themes. When describing our results only paraphrased quotes are included, sometimes with ethically fabricated information to maintain anonymity~\cite{markham2012fabrication}.  


\subsection{Results: Practitioner Discussions and Inquiries}
\label{sec:roles}

\begin{figure}[!ht]
\includegraphics[width=7cm]{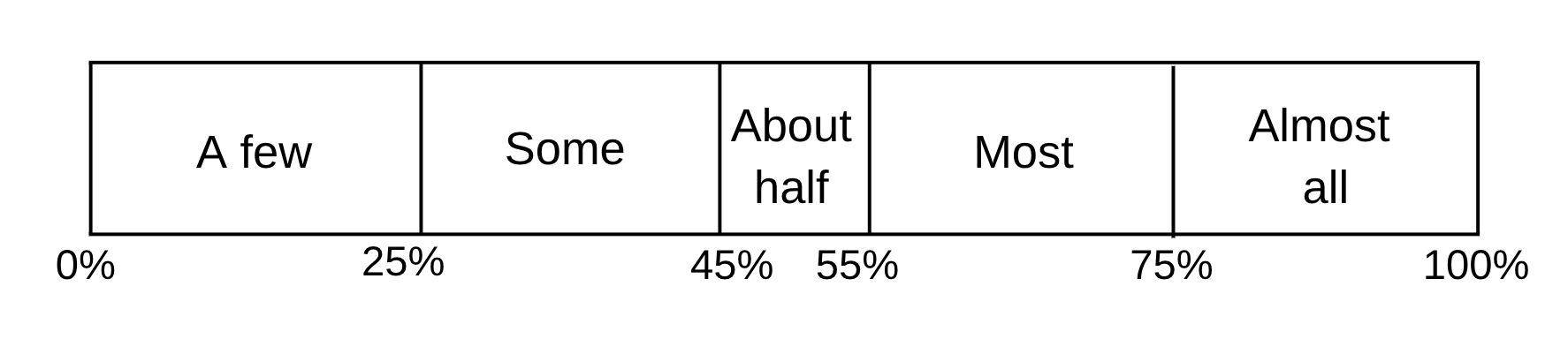}
\centering
\caption{The terminology used to represent the frequency of themes in qualitative research, related to inferred practitioner roles. This graphic was taken from ~\cite{kyi2023investigating}.}
\label{fig:frequencieswords}
This figure represents terminology used to describe the frequency of themes in qualitative research. `A few' refers to approximately 0 to 25\%, `Some' refers to approximately 25 to 45\%, `About half' refers to approximately 45 to 55\%, `Most' refers to approximately 55 to 75\%, and `Almost all' refers to 75 to 100\%.
\end{figure}

Due to the qualitative nature of this analysis, we do not provide quantified numbers of how many posts came from which role, and instead use terminology that has been used in previous qualitative research to describe relative frequencies (see Figure \ref{fig:frequencieswords})~\cite{kyi2023investigating,10.1145/3290605.3300764,10.1145/3313831.3376511}.

\subsubsection{Several practitioner roles are involved in GDPR compliance.}
Applying legitimate interests, and more broadly, the GDPR, is challenging for various types of practitioners. Often, posters explicitly indicated their role. However, other times we were able to infer their role based on the nature of the question. 
For example, we assumed that technical questions were posed by a developer, and client communication questions came from someone working within the marketing domain.
The majority of online GDPR compliance information tends to target developers, but our analysis showed that other roles are involved, or have an interest, in the application of legitimate interests, in line with recent work~\cite{stover2023website}. 


About half of the posts asked technical questions, therefore leading us to believe these posters might be \textit{developers}. There were some posts related to \textit{marketing} and some which seemed to come from \textit{business owners}. Additionally, a few posts were from other roles related to \textit{research}, \textit{human resources, business customers} (employees of a business that posed questions about how their business partner handles data), and other \textit{miscellaneous industries} (e.g., event planning, education, content creation, etc.). 

\subsubsection{GDPR accountability is difficult to determine.} 
\label{sec:accountability}
Deciding on the roles responsible for GDPR compliance is important for the concrete application of legitimate interests, since the lawful application of this legal basis has been doubted in previous research~\cite{kyi2023investigating,ferretti2014data}, therefore impacting the GDPR compliance for that company. We found that practitioners often queried about who in their company is responsible for GDPR compliance. 
%
Oftentimes, posters assumed it was the role of the developer to be the ``de facto'' data protection officer for their company, and some others believed it was the role of the owner of a given service. The following quote reveals why developers are often given the responsibility: \textit{``I'm a developer, not a lawyer, who is trying to comply with the GDPR and ePD as best as I can for my company. If we had more money, I would hire a lawyer to handle this for me, but we don't so I'm in charge of this.''} 
%
Discerning this role clearly is crucial since compliance obligations and liability depend on their accurate characterization (Recital 79 GDPR)~\cite{ico-guidance-controllers,EDPB-controller-2020}. Given the broad and flexible nature of legitimate interest as a legal basis~\cite{kamara2018understanding,ferretti2014data}, it is especially important to decide who is accountable for GDPR compliance within a team for ensuring the proper application of this legal basis.

\noindent \textbf{Determining accountability with third parties involved is challenging.}
%
In some cases, a poster mentioned that their service was not collecting any data, but third parties were, and therefore were confused about whether they (the first party service provider) would be accountable for GDPR data processing across their whole data supply chain. As discussed in previous literature~\cite{kollnig2021fait}, this problem reveals that practitioners are uncertain about who the data controller and processors were, who is accountable for compliance, and about the implications for joint controllership arrangements (as per Art. 26 GDPR).

\noindent \textbf{GDPR compliance relies on the compliance of others.}
%
%
Posts referred to scenarios where services relied on third party functionalities to function, therefore becoming difficult for practitioners to keep track of what data is collected and handled by each outsourced service. 
Some practitioners mentioned using non-compliant tools which impacted the compliance of their own service. Whether practitioners work alone, are part of a team, manage a development team, or are a service provider carrying out development services for third parties, it is essential to ensure that  personal data processing are sufficiently protected throughout the life-cycle of the project (according to the data protection by design principle (Art. 25 GDPR)). 
For example, a developer relied on WordPress to run their website; this same website used Google Fonts, which was considered non-compliant by a German Regional Court~\cite{google_fonts}, and therefore the developer was held accountable for a third party tool's processing and got fined. Such decisions echo recent case law on data controllership between a first and third party providers~\cite{C210/16,CaseC40/17,Planet49}. 


%
%


\label{sec:applyingLI}

\subsubsection{Choosing a legal basis for data processing can be confusing.}
\label{sec:LIvsothers}
Several posts aimed to clarify which legal basis practitioners should rely upon in their particular scenario. 
Often, legitimate interest was compared with consent (6(1)(a) GDPR), or contract necessity (6(1)(b)). 
In some cases, practitioners wondered if they could rely on both legitimate interest and consent, using legitimate interest as a fallback for users who could reject consent. This strategy might explain the prevalence of illegal cookie banners that use both consent and legitimate interest for the same processing purposes, a practice observed in a previous study~\cite{kyi2023investigating} and commented on by the European Data Protection Board (EDPB)~\cite{EDPBconsent2020}. 

\noindent \textbf{The legitimate interest legal ground is sometimes preferred over other legal grounds.}
\label{sec:whyLIusused}
Our analysis revealed that applying legitimate interest is an appealing legal ground for four many reasons.

First, 
this legal ground can be invoked as a last resource, such as when companies are unable to obtain prior consent, or in situations where they believed users would not consent, such as requesting users to consent to email marketing. The following quote shows how this legal ground may be abused to collect more data from users who did not consent: \textit{``Users who don't consent to advertising are a problem because my app relies on ads, therefore I'll lose revenue if ads aren't displayed. Is there a way to make it required for users to consent? Otherwise I won't be able to offer my service anymore.'' } This observation supports legal speculation that legitimate interests could potentially be abused as a loophole to collect more data~\cite{kamara2018understanding,ferretti2014data}. 
%

Second, legitimate interests might be used for the mere convenience of the practitioner. For example, when a system is already collecting user data before consent can be given, or when a user does not have access to a system to give consent, applying the legitimate interest legal ground is more convenient compared to asking for retroactive consent. 

Third, legitimate interest may be used because other legal grounds for data collection do not apply to their scenario. 
Our analysis revealed that there were a variety of potential legitimate interest use cases practitioners posted about which reveals the benefits of having a broad and flexible legal ground.

Fourth, many posts held the assumption that legitimate interest may be applied to retain data in a scenario where users exercise the \textit{``Right to be Forgotten''} (Art. 17 GDPR). 
This has practical implications because controllers can lose money if data is deleted, and some companies may face challenges locating the user's data upon an account deletion request, according to our analysis. 

\subsubsection{Questions from practitioners indicate knowledge, but also misunderstandings of the GDPR}
\label{sec:devsconfused}
Some practitioners mentioned some legal arguments in their questions, and other cases revealed that practitioners already had previous experience with the GDPR, but could not get concrete granular information needed for their particular situation. Various examples in our dataset demonstrate that practitioners have a sincere intention to process personal data in a compliant manner: \textit{``I've read the entire GDPR but I'm still confused.''}

Trying to comply with the GDPR is made more difficult when it comes to the legitimate interest legal ground, as these are more broad and flexible in their application, with little guidance on how they should be used. While other legal bases such as consent are well-established with the advent of cookie banners and the general understanding that companies need to request user consent in order to collect and process data, there is less knowledge about the legitimate interest legal ground~\cite{kyi2023investigating}. However, we find that developers (and those applying the GDPR) were often confused by their \textit{own} service's data processing. 
The examples below illustrate commons points of confusion.


\noindent \textbf{Practitioners are confused about data collection and tracking.}
Some practitioners were unsure of what data was being tracked, what data was stored in which cookie(s), what happens to user data after it is collected, and what data third parties have access to: 
\textit{``I'm using a third-party tool for the app I've developed, but I don't know which server users will be routed to, or what country the server is in.''}

Some practitioners were unsure about data collection implications in other scenarios, such as the data collected if a user was registered or not, and what to do if they receive a user account deletion request. There was also some confusion about what practitioners and organizations should do with user data, and whether they could rely on legitimate interest to keep user data. This highlights the difficulty of determining which legal basis and purpose should be applied to certain data. 

\noindent \textbf{There are some misunderstandings around consent.}
Practitioners were sometimes unable to tell if a user consented to data processing or not, and mentioned that the law does not explain how to \emph{use} consent, only how to \emph{get} consent. In fact, the most viewed posts in our sample referred to questions about whether the purposes of certain tracking technologies were \emph{strictly necessary} for a service to work, and hence, whether they could be exempted from consent (as per Art. 5(3) ePrivacy Directive). 

However, the European Data Protection Board (EDPB) and DPA guidelines explain with detail when trackers are to be considered essential for services to function, and as such, exempted from consent~\cite{EDPBconsent2020,29WP-4-12-CookieExemption,ICO-Guid-19}. Some posters faced doubts about how consent impacts their data practices, such as what services can be offered and what data should then be collected if users reject consent.

Previous work has shown that the legitimate interest ground is commonly seen in cookie banners, being treated similarly to consent~\cite{kyi2023investigating,matte2020purposes}. Therefore, the challenges practitioners face in distinguishing whether users have provided valid consent indicate that companies may struggle to manage and categorize the data collected from users and align it with the appropriate legal basis. 

\noindent \textbf{Anonymization and pseudonymization are misunderstood.}
\label{sec:anon}
%
Oftentimes, it was mentioned that true anonymization is difficult, and just because data is hashed, it does not mean it can evade GDPR requirements. However, there was discussion about whether anonymization and pseudonymization would be sufficient to protect user's personal data, and whether the GDPR applies to anonymized and pseudonymized data: \textit{``I believe the GDPR wouldn't apply if I properly anonymize the data because it wouldn't count as personal information.''}

\noindent \textbf{What counts as personal data is misunderstood.}
\label{sec:personaldata}
We found that most held conflicting beliefs of what counted as personal data. 
Some believed that data is not personal if it only identifies a device instead of an individual person, while others correctly understood that the GDPR has a broad concept of what counts as \textit{personal data} and what is an \emph{identifiable person}.  
%

A few practitioners mentioned personal data as being anything from nicknames and usernames, any demographic data, photos and videos of others, and user emails. Data gathered from users, such as medical data and Google searches could also count as personal data, according to some posts. 
These disagreements and misunderstandings highlight how opaque and complex industry data practices can be for developers and service owners, and the potential consequences for GDPR compliance that might occur when even those building the services are unsure of data processing requirements. 
%



\section{Analysis 2: Legal Assessment of GDPR Compliance Information in Online Forums} 
\label{sec:results:legalsoundness}

In this section we analyzed 94 \emph{accepted answers} to the posts collected from Law Stack Exchange. 
We use the term ``poster'' to refer to the practitioner who originally posted a query, and ``commentator'' to the one providing an answer. 


\noindent \textbf{Dataset.} We filtered our Law Stack Exchange dataset used in the first round of analysis to include only posts that were marked by the original poster as ``Accepted''~\cite{accepted}. ``Accepted answers'' are answers where the original poster of a question marked a certain answer as the one which most solved their problem~\cite{accepted}. We deliberately focused on these answers because other forum users with similar use cases might assume they contain ``correct'' information to solve their problems~\cite{tahaei2022privacy}, and might act according to the answer(s) received. This yielded 94 answers. 

\noindent \textbf{Annotation.} One co-author, a legal scholar specialized in EU data protection law and with over five years of experience, analyzed the answers. 
%

The evaluation of the \textit{Accepted} forum answers consisted upon the following criteria: 
i) \emph{Legal soundness}: whether answers gave \textit{sound, partly sound,} or \textit{not sound} legal information to the post's question(s); and ii) \emph{Completeness}: whether answers gave \textit{complete} or \textit{incomplete} legal information to the question posted.
 
To analyze the legal soundness of the answers, it was necessary to resort to legislation, judicial and regulatory decisions, and guidelines from the EDPB or from DPAs to confirm whether the provenance of some arguments in the given answers hold. As some questions included technical terminology, it was necessary to further understand the background technical scenario to fully assess the content of the answers.

\subsection{Results: Legal Soundness}
In our analysis, based on the information provided in the post, we found that the majority (73.4\%) out of 94 accepted answers were deemed to be legally sound; 20.2\% were partly legally sound, and only 6.4\% were not legally sound. Out of these 69 legally sound answers, 8.9\% were incomplete, meaning they were missing information that would be useful for practitioners to know when applying legitimate interests. 

We note that the majority of answers appeared legally sound due to the low complexity of the questions, e.g., \emph{``Should the session end if users reject cookies?''}, \emph{``Do low-effort cookie banners comply with the GDPR?''}, \textit{``Does tracking IP addresses count as processing personal data?''}, or \textit{``Will opting in users into phone push notifications count as a GDPR violation?''}. In these situations, legal soundness would apply regardless of the poster's jurisdiction or the type of organization they work at due to the general nature of these questions. 
%

\subsubsection{Completeness}\hfill

The \textit{Completeness} of forum answers is due to the posts and answers to these posts. Some posts contained several sub-questions or were very broad, and some answers involved mentioning other legal sources and legal terminology, which contributed to the legal \textit{Completeness} of an answer.

\noindent 
\textbf{It is difficult to comprehensively answer several questions.}
Some posters included several sub-questions in a single post, 
similar to: \textit{``Is this legal? Do I have any say in it? Can I refuse to use this service? What other data could they have access to?''}
In such cases, we observed that, even if the answer appeared to be sound, some subquestions tended to remain unanswered.

\noindent \textbf{General questions trigger general answers.} 
In response to very general questions (\textit{``Do GDPR rules apply to my case?''}), often several hypotheses were raised, especially when the concrete purpose for processing personal data was unclear: \emph{``It's not likely I can provide a correct answer because it depends on what you're trying to achieve.''} 
The answers then remained general even if a question was edited several times with more factual information added by the poster. 
The presence of such general answers might lead to poor compliance practices because it is important to discern which GDPR rules the poster is concerned about in order to avoid a catch-all response.

\noindent \textbf{Answers sometimes include citations of legal sources and national laws.}
GDPR legal provisions and even Recitals were convened in the answers. Notably, excerpts from case law of the Court of Justice of the EU were used (eg.,~\cite{Breyer2016,Nowak2017,CaseC40/17,C210/16}), as well as guidelines from the EDPB (mostly on consent~\cite{EDPBconsent2020,29WP-4-12-CookieExemption,Transparency29WP} and behavioural targeting~\cite{opinionads}), and several UK DPA guidelines. 
When some posts questioned the applicability of certain national laws, 
answers provided excerpts of the consulted legal provisions supporting their explanations. 
%

However, in our sample, no regulatory decisions from DPAs were cited (which include concrete cases that could inform several questions and answers closely related to the posts), even if there exists a publicly available online repository with relevant DPA decisions and more general GDPR insights that can be shared across Europe~\cite{GDPRhub}.

\noindent \textbf{The ePrivacy Directive is rarely cited.}
Some posters asked whether consent was required for tracking technologies such as browser fingerprinting, third party website resources (e.g., Google Fonts, embedding GitHub Gists, etc.), and analytics. Answers mostly ignored the applicability of the ePD~\cite{ePD-09} to such cases, 
potentially because of the incorrect conviction that only the presence of cookies triggers the application of this directive~\cite{EDPB-guidelines19B}. 
The ePD was furthermore mentioned neither when some commentators said that Google Analytics could count as a legitimate interest, nor in answers concerning assigning a unique ID to a device for tracking location data. 

\noindent \textbf{Answers use imprecise legal terminology.}
While most answers were generally legally sound, we noted imprecision in the adopted terminology. Examples include the use of the term ``personal identifiable information (PII)'' from American data protection law instead of ``personal data'' used in the EU, references to ``permission to use the data'' instead of ``consent'', stating that ``permission must be clear and positive'' to indicate unambiguous consent, or even inaccurate explanations of the three-tiered balancing test~\cite{EDPB-6-14} of the legitimate interest legal basis. 


\subsubsection{Soundness}\hfill

Here, we discuss common areas where commentators did not give legally sound information. 

\noindent \textbf{Incorrect legitimate interest-based purposes are sometimes suggested.}
Several commentators asserted that posters could use the legitimate interest legal ground for purposes such as the use of Google Ads, online exam proctoring, or personalized advertising.

While the legal basis of legitimate interest is open-ended (with a broad and unspecific scope), several regulatory decisions and EDPB guidelines determined that personalized advertising purposes cannot rely on legitimate interest grounds~\cite{WP29-Ads-10,BelgianDPAdecision-IAB2022,EDPB-6-14,matte2020purposes,ecj_case_2023}. Moreover, some legal scholars argued that legitimate interests should not be used as a ground to collect data by higher education institutions for proctoring~\cite{fouad2018lawful}. This reveals that there is a mismatch between the legal field and some commentators' conceptions of legitimate interests, which may result in this legal ground being misused if posters were to follow this legal information.


\noindent \textbf{Legitimate interests are incorrectly invoked under the \textit{``Right to be Forgotten''}.}
%
Several posters wanted to know if they could deny a request for account deletion from their users under legitimate interests. They moreover wondered whether they could keep data based on legitimate interests in cases where relevant data records could not be located. 
%
%
%
Answers related to these posts mentioned the right to erasure, but did not account for all the legal grounds for this right to be exercised (under Article 17(1)), or its exceptions (prescribed in Article 17(3) and Recital 65) that could be potentially applicable to override the deletion request. We argue that incomplete answers about erasure obligations are mostly due to the difficulty of operationalizing such grounds and exceptions in practice~\cite{ausloos}.

\noindent \textbf{Legitimate interests are invoked without considering a balancing test.}
Several answers suggested that posters could deny data deletion requests on the ground of ``overriding legitimate grounds for the processing'' (Article 17(1)(c)) and provided several concrete examples, such as keeping a revision history in the interest of security, 
or storing a certificate after an account is deleted.
However, 
these suggestions did not duly indicate that the data controller bears the burden of proof, meaning to demonstrate whether those recommended \emph{overriding} legitimate grounds do indeed exist~\cite{EDPB-5-19,kuner2021eu,judgment}. If a controller fails to demonstrate the existence of overriding legitimate grounds, the data subject is entitled to have their data deletion request executed. 

Sound answers regarding this issue of the balancing test ought to account for several factors~\cite{NorwegianDPAdecision-MetaLI2023}. 
Such factors include how compelling the legitimate interest of the controller is; the nature and source of the legitimate interests; the degree of impact on the interests, rights and freedoms of the data subjects; the nature of the data; the way that data are being processed; the source and accessibility of the data; the reasonable expectations of the data subject; the status of the data controller and data subject as well as what safeguards are in place beyond the minimum required by the GDPR.

\noindent \textbf{Legitimate interests are incorrectly invoked in the context of publicly available data.}
In response to posts asking about whether they can use publicly available data for their commercial purposes, some commentators asserted that scraping data publicly available on the web could be based on legitimate interests and it would suffice to disclose the practice in a privacy policy on a relevant website, as per Article 14 GDPR. 
The reuse of publicly available online data for commercial purposes requires user consent in most data processing scenarios~\cite{CNIL_Ouverture_Reutilisation_Donnees,EDPB_Guidelines_LegitimateInterest_2024,FrenchDPA-decision-scraping,IrishDPA-decision-scraping}. 
The legitimate interest basis can only in theory justify reuse of this data if the the three-tier test is fulfilled
and most importantly, considers large-scale data collection,
user's reasonable expectations of their data being reused , and inclusion of sensitive data, requirements that ultimately impede the use of this legal grounding. 


The answers also did not account for potential sensitive categories of data, nor the obligation to directly and \emph{actively} inform the involved data subjects about their rights (e.g., to object to legitimate interests), the source of the scraped data, as well as the mandatory information required by Article 14(1-3) GDPR~\cite{Transparency29WP,decisionPolishDPA-webscraping}. 
Yet, some answers acknowledged that the public nature of the scraped data might be factored in the legitimate interest test: \textit{``The data subject making this info public is a factor in the legitimate interest balancing test, but it depends on how you use the data and if they can expect their data to be used this way.''}

\noindent \textbf{Consent requirements are interpreted incorrectly.}
Some answers did not regard the legal requirements for consent requests~\cite{EDPBconsent2020}, in particular the \textit{prior, unambiguous} and \textit{revocable} requirements. One commentator assured that \emph{``An email with an opt-out function should be compliant''} as it offers a way to withdraw consent that is as easy as it is to give consent (as per Article 7(3), Recital 42 GDPR).
The legal requirement for an \textit{unambiguous consent request} was not always clearly explained, despite several EDPB guidelines clarifying that consent should be given by a clear affirmative action from users (Article 4(11), Recital 32)~\cite{Sant-etal-20-TechReg} and as such, silence, inactivity, or other assumptions of related passive actions are illegal. Lastly, it is not always understood what the ``strict necessity exemption'' of Article 5(3) ePD entails and which purposes should be considered essential and necessary to provide a service explicitly requested by a user.
%

\noindent \textbf{The \textit{Contract Necessity} legal basis is generally not mentioned.}
As mentioned in Section~\ref{sec:background}, contract necessity is a lawful basis for processing personal data ``necessary for the performance of a contract'' in which the data subject is involved in 
(Article 6(1)(b), (Recital 65))~\cite{contract}.
However, even if an implicit contractual relationship was mentioned in the main question of the post, this legal basis was mostly not discussed in the answers, and instead posters were encouraged to rely on consent or legitimate interest as a legal basis.

\noindent \textbf{The concept of personal data is not always properly understood.}
Understanding the concepts of ``personal data'' and ``processing'' is essential for the compliant development of any application reliant upon user data, and the GDPR applies to any data that is identifiable. Therefore, misunderstanding what personal data is can result in incorrect applications of the GDPR.
However, we observed that some answers provided an incorrect definition or erroneous examples of applications of the concept of personal data, as per Article 4(1). For instance, some answers did not consider online identifiers, like IP addresses, as potentially identifiable data~\cite{Breyer2016,EDPB-4-07}: \textit{``It's personal data if they're a registered user and you can identify the data subject, but if they aren't registered and you only collect their IP address, it's not personal data''}. 

\section{Discussion}
\label{sec:discussion}

We analyzed 319 online forum posts from practitioners who asked questions relating to the legitimate interest legal basis, along with 94 \textit{accepted} answers. We discuss the implications of our results regarding the application of the legal ground of legitimate interests and GDPR compliance. 


\noindent \textbf{Posters are motivated to comply but still need guidance in applying legitimate interests in practical settings.}
Those who do not care about compliance or are malicious in their compliance are unlikely to seek legal information in the first place. However, of those who did post a question, it does not seem like \textit{most} posters are actively trying to circumvent the law based on the amount of posts seeking legal information, and the amount of views posts had. 
Rather, our analysis suggests there is a wide range of potential areas of confusion and doubt amongst practitioners, therefore showing that practitioners require more guidance when applying legitimate interests and the GDPR in practice. 

\noindent \textbf{Several factors increase the chances of non-compliance.} Our findings suggest that non-compliance is often due to various factors. First, practitioners may misunderstand the complex legal language and open-textured concepts (like \emph{strict necessity} or \emph{balancing test}, and the multitude of stipulations in the GDPR and the ePD). Second, there is a lack of legal training amongst practitioners in charge of developing services that require GDPR compliance. 
Third, the reliance on third-party tools which are not always GDPR-compliant creates a domino effect of non-compliance for services using these tools. 
Lastly, it is difficult to find reliable information online; some posters mentioned seeing contradicting information, or not finding information specific to their scenario. All combined, these issues make GDPR compliance a difficult feat, thereby potentially resulting in privacy implications for end-users of a service. 

\noindent \textbf{Legal compliance is a multi-stakeholder issue.} 
Our study shows that posters asking about the GDPR potentially come from a variety of roles, such as developers, marketers, business owners, human resources, etc. 
We consider it to be important that DPAs, the EDPB, and other relevant organizations provide more concrete guidelines and recommendations tailored to practitioners, while considering different roles. It should be more clear when one is a data controller, processor or within a joint controller relationship. 


\noindent \textbf{There are several potential legitimate interest use cases.} Our analysis showed the
breadth of potential cases under which legitimate interest may apply, 
but also highlighted how legitimate interests can potentially be misused and abused by practitioners, especially in situations where consent is difficult to obtain. Often, legitimate interests were shown to be a preferred legal ground, and commentators did not often mention other legal bases posters could use, nor did they mention the proper provisions needed to invoke legitimate interests. 

\noindent \textbf{Time and investment from commentators might give credibility.} In our analysis, most \textit{Accepted} answers were considered as legally sound based on the information provided. Comprehensive answers (with case law notes, excerpts of guidelines, and references to national laws) denote time and attention from commentators. 
Unlike Stack Overflow, which can be used in job applications to display one's technical knowledge and abilities~\cite{stackoverflowjobs}, those answering Law Stack Exchange questions did not generally specify if they had a legal background, nor was it clear what the incentive is for commentators to participate and give thorough answers for free. 

\noindent \textbf{Dedicated resources exist but are not referred to in forums.}
Online forums may be preferred by practitioners to find legal information because of the granularity of questions and answers which relate to applying the GDPR for a specific scenario. 
The legal sources that were cited in the answers do not mention already-available online resources for developers mentioned in Section~\ref{sec:onlinelegalinformation}, but instead referred to DPA guidelines. 
%
The granularity of concrete scenarios that developers face seem to have inspired the creation of the French Data Protection Authority's (CNIL) developer guide~\cite{CNIL_GDPR_Developers_Guide}, which is an example of how the law can be operationalized to more concrete settings. Interestingly, this resource was not mentioned in posts or comments we analyzed. 
%
%



\subsection{Implications for Legal Compliance}
\label{sec:implications}


\textbf{Online forums have an implicit role in data protection.} 
Based on the number of posts relating to GDPR questions, and the number of views and comments under these posts, it suggests that some practitioners are turning to these forums to find information that is most applicable to their specific scenario to better comply with the GDPR. Therefore, this indicates that there is a strong case for considering the role of online forums for GDPR compliance. Therefore, DPAs could provide more applicable scenario-based guidance and specific technical requirements so that practitioners can better understand how to properly apply the GDPR in practical contexts.

Resorting to these forums may indicate a lack of legal resources for smaller and medium-sized companies who in turn might not have in-house legal department at their disposal to discuss GDPR compliance issues. 
We posit that the status and the social responsibility of online legal forums merit attention, and call for discussions on the features that should be added or removed to make the sourced information more helpful, accurate, complete, and shaped to the practitioners' needs. 

Based on our findings, we present several suggestions for how to better design online legal forums to help practitioners apply and understand the GDPR correctly. 
Forums could include, for example: 
i) practical resources to assist practitioners in setting up law-abiding services according to their sector of activity, especially addressing scenarios where GDPR legal provisions are vague and use open textured concepts, 
ii) frequently asked questions (FAQs) on the legal bases and purposes used, 
iii) a large typology of examples of data that could be considered personal, and 
iv) implementable scenarios (per sector), 
v) reconcile contradictions between posts on the same topic
vi) regular check ups of answers to consider possible updates of legislation and new DPA guidelines.

\noindent  \textbf{Disclaimers of incomplete or incorrect answers.} 
It is difficult to establish the legal expertise of those answering the posts; some mentioned having legal expertise in their profile pages (sometimes with a link to their personal webpage, or using a nickname which is also used on other social media), while the majority did not refer to their profession.
Even though some more straightforward answers were sound (to the point of citing legal sources) and recommended examples of best practices, commentators usually included disclaimers that posters should get legal information from a lawyer for a rigorous determination of their case. 
%

Even if a given answer is not correct, a practitioner is not in a position to hold a commentator liable when incorporating such legal information into their services/products. Law Stack Exchange itself states that it ``is for educational purposes only and is not a substitute for individualized information from a qualified legal practitioner. Communications on Law Stack Exchange are not privileged communications and do not create an attorney-client relationship''~\cite{lawse}. Such a disclaimer might exonerate the forum and commentators from any legal responsibility regarding the answers given, though this might not be sufficiently understandable nor visible for users seeking legal information and receiving detailed feedback. Therefore, these forums may act as a filter to help posters discern the level of legal complexity and whether there is a need to get help from a lawyer, and the level of risk they are taking if they do not consult with one.



\subsection{Guidelines Towards Writing Forum Posts for Legally-Sound Answers}
\label{sec:guidelines}
GDPR compliance relies on the efforts of several stakeholders. As such, those involved in GDPR compliance for an organization are in charge of a large and important task that requires not only understanding legal documents and the requirements for compliance, but also applying the law into practice. 

Our results suggest that there are several areas of confusion practitioners may have about applying legitimate interests into practice, and there are several ways that legal compliance can go wrong if given incorrect advice on a forum. It is also possible that in the near future, more practitioners and users may start turning to AI chatbots for legal information~\cite{law_gpt}, which are also trained on online forum data~\cite{hamalainen2023evaluating,sanh2021multitask}. 

Therefore, it is important that online forums provide correct legal information for current and future uses of this information for legal compliance. Our analysis found that most commentators take time and care in crafting their responses, but legal soundness is also reliant on other factors, such as a poster's jurisdiction, practices and industry. In this section, we present three recommendations to ensure that posters can receive responses that are as legally sound as possible, which online legal forums could include in their rules and guidelines for posters.

First, posters should include the jurisdiction of their organization, and the jurisdiction of their user base. The jurisdiction should specify the country at stake, since different European Member States may have different national law specifications. 
Second, posters should also mention i) the industry their organization is involved in, especially if it deals with sensitive data, such as health information, so that posters know of special safeguards they need to adopt with users' data, ii) processing purposes, and iii) their concrete role within data processing activities. 
Third, commentators should provide a disclaimer about the limitations of the concrete answer and add the fact that the dedicated legal information should neither bind the poster nor be construed or interpreted as legal advice.

\subsection{Limitations.} 
\textbf{Doubts of generalizable results to other forums and legal bases.} The first limitation is that of generalizability. Since we only checked the accepted answers on Law Stack Exchange relating to the legitimate interest legal basis, we are not sure how generalizable our results are to other forums or legal bases. Additionally, we have a sample bias as we only analyzed online forums, therefore it is possible that compliance with GDPR's legitimate interests might be more straightforward for more practitioners. 



\noindent \noindent \textbf{Our legal analysis of answers is limited to information provided.} The legal analysis in Section~\ref{sec:results:legalsoundness} was merely confined to the facts and bounded to the scenarios described by the poster, 
and to the interpretation sustained by the commentator.
Accordingly, legal ambiguities within these answers were observed due to the limited knowledge and context circumscribed by each post, such as the type of organization posters work at, whether posters were actually data controllers (processors, sub-processors, or none), the accuracy of the processing purposes mentioned, the nature and sensitivity of the data collected, the exact way data is processed, the source and accessibility of the data, or the concrete jurisdiction or national law the poster refers to, among other circumstances. 

As such, our analysis on the correctness and completeness of the answers is thus limited and constrained to the reported posted facts and interpretations given by the commentators.  
Therefore, we acknowledge that there is a margin of doubt while labeling answers as ``legally sound'', ``not legally sound'' or ``incomplete''. We emphasize that only a judicial assessment that requires more specific fact-finding of each respective question and answer could render a final appraisal of such analysis and provide definitive certainty.

Due to these limitations, we therefore proposed guidelines in Section~\ref{sec:guidelines} for framing legal compliance questions in our Discussion section to ensure more legally sound responses to posters' questions in the future. 

\subsubsection{Future directions}
\label{sec:future}
Based on the observed usage of online forums for legal information in our study, we can foresee LLMs being used in the future for quick legal compliance information. Using crowdsourced data for legal compliance information, whether from online forums or LLMs, can have a wide range of legal and ethical implications which are yet to be explored~\cite{veale2018algorithms}. To mitigate the risks that come with using crowdsourced data for legal compliance information, more measures must be taken to ensure that high-quality, legally-sound information is being shared in online forums and LLMs. As such, we believe future work in this space could investigate how practitioners use LLMs for legal information, and further discuss the ethical and legal challenges and protections available for users of these services.  

As GDPR compliance involved several stakeholders not only limited to those in law, but also those in tech, we suggest more collaborations between human-computer interaction researchers and the legal community to better understand the human factors of legal compliance and create solutions for practitioners involved. 




\section{Conclusion}
\label{sec:conclusion}

In this paper, we investigated the practical challenges of legal compliance for practitioners by analyzing posts and answers from online forums (Reddit, Stack Overflow, and Law Stack Exchange), using legitimate interest as a case study. Our two-part analysis indicated that: i) finding out how to be compliant is difficult mostly for small and medium-sized companies, ii) practitioners seek information for their specific scenarios, and iii) legal compliance information from \textit{Accepted} Law Stack Exchange answers are generally legally sound, but sometimes incomplete, with the mentioned limitations of our analysis. 

We believe that legitimate interests are prone to being misused due to the broad and flexible nature, leading practitioners to misunderstand their concrete applications. In light of this, the human-computer interaction and legal communities should consider the role of online forums in the data protection landscape, and pay more attention to the human factors of legal compliance to give more actionable and scenario-specific guidelines to foster data protection by design and by default. 





\bibliographystyle{plain}
\bibliography{sample-base}

\begin{thebibliography}{10}

\bibitem{WP29-Ads-10}
Opinion 2/2010 on online behavioural advertising, 2010.
\newblock \url{https://ec.europa.eu/justice/article-29/documentation/opinion-recommendation/files/2010/wp171_en.pdf}.

\bibitem{Breyer2016}
{Case 582/14 – Patrick Breyer v Germany}.
\newblock Court of Justice of the European Union ECLI:EU:C:2016:779, 2016.

\bibitem{Nowak2017}
{Case C-434/16 - Peter Nowak v Data Protection Commissioner}.
\newblock Court of Justice of the European Union ECLI:EU:C:2017:994, 2017.

\bibitem{GDPR4Devs}
{GDPR for Developers}.
\newblock \url{https://gdpr4devs.com/}, 2025.

\bibitem{acar2016you}
Yasemin Acar, Michael Backes, Sascha Fahl, Doowon Kim, Michelle~L Mazurek, and Christian Stransky.
\newblock You get where you're looking for: The impact of information sources on code security.
\newblock In {\em 2016 IEEE Symposium on Security and Privacy (SP)}, pages 289--305. IEEE, 2016.

\bibitem{BelgianDPAdecision-IAB2022}
{Decision on the merits 21/2022 of 2 February 2022 Complaint relating to Transparency \& Consent Framework}, 2022.

\bibitem{law_gpt}
Kathryn Armstrong.
\newblock Chatgpt: Us lawyer admits using ai for case research, 2023.

\bibitem{attribution}
Jeff Atwood.
\newblock Attribution required, 2009.

\bibitem{ausloos}
Jef Ausloos.
\newblock {\em The Right to Erasure in EU Data Protection Law}.
\newblock Oxford University Press, Oxford, UK, 2020.

\bibitem{bacchelli2012harnessing}
Alberto Bacchelli, Luca Ponzanelli, and Michele Lanza.
\newblock Harnessing stack overflow for the ide.
\newblock In {\em 2012 Third International Workshop on Recommendation Systems for Software Engineering (RSSE)}, pages 26--30. IEEE, 2012.

\bibitem{barakat2022community}
Hanna Barakat and Elissa~M Redmiles.
\newblock Community under surveillance: Impacts of marginalization on an online labor forum.
\newblock In {\em Proceedings of the International AAAI Conference on Web and Social Media}, volume~16, pages 12--21, 2022.

\bibitem{barua2014developers}
Anton Barua, Stephen~W Thomas, and Ahmed~E Hassan.
\newblock What are developers talking about? an analysis of topics and trends in stack overflow.
\newblock {\em Empirical Software Engineering}, 19(3):619--654, 2014.

\bibitem{beadle2025sok}
Kyle Beadle, Kieron~Ivy Turk, Aliai Eusebi, Mindy Tran, Marilyne Ordekian, Enrico Mariconti, Yixin Zou, and Marie Vasek.
\newblock Sok: A privacy framework for security research using social media data.
\newblock In {\em 2025 IEEE Symposium on Security and Privacy (SP)}, pages 1178--1196. IEEE, 2025.

\bibitem{bednar2019engineering}
Kathrin Bednar, Sarah Spiekermann, and Marc Langheinrich.
\newblock Engineering privacy by design: Are engineers ready to live up to the challenge?
\newblock {\em The Information Society}, 35(3):122--142, 2019.

\bibitem{aircanada}
Ashley Belanger.
\newblock Air canada has to honor a refund policy its chatbot made up, 2024.
\newblock \url{https://www.wired.com/story/air-canada-chatbot-refund-policy/}.

\bibitem{EDPB-4-07}
European Data~Protection Board.
\newblock Opinion 4/2007 on the concept of personal data ({WP} 136), adopted on 20.06.2007, 2007.

\bibitem{contract}
European Data~Protection Board.
\newblock Guidelines 2/2019 on the processing of personal data under article 6(1)(b) gdpr in the context of the provision of online services to data subjects, 2019.

\bibitem{EDPB-5-19}
European Data~Protection Board.
\newblock Guidelines 5/2019 on the criteria of the right to be forgotten in the search engines cases under the gdpr (part 1), 2020.

\bibitem{braun2006using}
Virginia Braun and Victoria Clarke.
\newblock Using thematic analysis in psychology.
\newblock {\em Qualitative research in Psychology}, 3(2):77--101, 2006.

\bibitem{bygrave2017data}
Lee~A Bygrave.
\newblock Data protection by design and by default: Deciphering the eu’s legislative requirements.
\newblock {\em Oslo Law Review}, 4(2):105--120, 2017.

\bibitem{FrenchDPA-decision-scraping}
CNIL.
\newblock Délibération san-2020-018, 2020.

\bibitem{irishdpa}
Data~Protection Commission.
\newblock Data protection commission announces conclusion of two inquiries into meta ireland, 2023.

\bibitem{CNIL_GDPR_Developers_Guide}
{Commission Nationale de l'Informatique et des Libertés (CNIL)}.
\newblock {GDPR Developer's Guide}.
\newblock \url{https://www.cnil.fr/en/gdpr-developers-guide}, June 2020.

\bibitem{CNIL_Ouverture_Reutilisation_Donnees}
{Commission Nationale de l'Informatique et des Libertés (CNIL)}.
\newblock {Ouverture et réutilisation de données personnelles sur Internet: la CNIL publie ses recommandations}.
\newblock \url{https://www.cnil.fr/fr/ouverture-et-reutilisation-de-donnees-personnelles-sur-internet-la-cnil-publie-ses-recommandations}, June 2024.

\bibitem{da2025llms}
Leuson Da~Silva, Jordan Samhi, and Foutse Khomh.
\newblock Llms and stack overflow discussions: Reliability, impact, and challenges.
\newblock {\em Journal of Systems and Software}, page 112541, 2025.

\bibitem{IrishDPA-decision-scraping}
Irish DPA.
\newblock Data protection commission reference: In-21-4-2 in the matter of meta platforms ireland ltd. (formerly facebook ireland ltd.), 2022.

\bibitem{decisionPolishDPA-webscraping}
Polish DPA.
\newblock Decision zspr.421.3.2018, 2019.

\bibitem{dym2020ethical}
Brianna Dym and Casey Fiesler.
\newblock Ethical and privacy considerations for research using online fandom data.
\newblock {\em Transformative Works and Cultures}, 33, 2020.

\bibitem{ecj_case_2023}
{ECJ}.
\newblock Case {C}-252/21: {Request} for a preliminary ruling from the {Oberlandesgericht} {Düsseldorf} ({Germany}) lodged on 22 {April} 2021 — {Facebook} {Inc}. and {Others} v {Bundeskartellamt}, 2023.

\bibitem{EDPB-guidelines19B}
European Data Protection~Board (EDPB).
\newblock Opinion 5/2019 on the interplay between the eprivacy directive and the gdpr, in particular regarding the competence, tasks and powers of data protection authorities, 2019.

\bibitem{10.1145/3290605.3300764}
Pardis Emami-Naeini, Henry Dixon, Yuvraj Agarwal, and Lorrie~Faith Cranor.
\newblock Exploring how privacy and security factor into iot device purchase behavior.
\newblock In {\em Proceedings of the 2019 CHI Conference on Human Factors in Computing Systems}, CHI '19, page 1–12, New York, NY, USA, 2019. Association for Computing Machinery.

\bibitem{EUdataregulations2018}
{European Commission}.
\newblock {2018 {R}eform of {EU} data protection rules}.
\newblock Available at \url{https://ec.europa.eu/commission/sites/beta-political/files/data-protection-factsheet-changes_en.pdf}, 2018.

\bibitem{EDPB-controller-2020}
{European Data Protection Board}.
\newblock {Guidelines 07/2020 on the concepts of controller and processor in the GDPR Version 1.0}, 2020.
\newblock \url{https://edpb.europa.eu/our-work-tools/public-consultations-art-704/2020/guidelines-072020-concepts-controller-and-processor_en}.

\bibitem{EDPB-6-14}
{European Data Protection Board (EDPB)}.
\newblock Opinion 06/2014 on the notion of legitimate interests of the data controller under article 7 of directive 95/46/ec (wp 217), 2014.

\bibitem{EDPBconsent2020}
{European Data Protection Board (EDPB)}.
\newblock Guidelines 05/2020 on consent under regulation 2016/679, 2020.

\bibitem{EDPB_SME_Data_Protection_Guide}
{European Data Protection Board (EDPB)}.
\newblock {SME Data Protection Guide}.
\newblock \url{https://www.edpb.europa.eu/sme-data-protection-guide/home_en}, April 2023.

\bibitem{EDPB_Guidelines_LegitimateInterest_2024}
{European Data Protection Board (EDPB)}.
\newblock {Guidelines 1/2024 on processing of personal data based on Article 6(1)(f) GDPR}.
\newblock \url{https://www.edpb.europa.eu/system/files/2024-10/edpb_guidelines_202401_legitimateinterest_en.pdf}, October 2024.

\bibitem{Transparency29WP}
Law~Stack Exchange.
\newblock General disclaimer, 2023.

\bibitem{accepted}
Stack Exchange.
\newblock How does accepting an answer work?, 2019.

\bibitem{stackexch}
Stack Exchange.
\newblock Academic papers using stack exchange data, 2022.

\bibitem{ferretti2014data}
Federico Ferretti.
\newblock Data protection and the legitimate interest of data controllers: Much ado about nothing or the winter of rights?
\newblock {\em Common Market Law Review}, 51(3), 2014.

\bibitem{finck2021reviving}
Michele Finck and Asia~J Biega.
\newblock Reviving purpose limitation and data minimisation in data-driven systems.
\newblock {\em Technology and Regulation}, 2021:44--61, 2021.

\bibitem{fischer2017stack}
Felix Fischer, Konstantin B{\"o}ttinger, Huang Xiao, Christian Stransky, Yasemin Acar, Michael Backes, and Sascha Fahl.
\newblock Stack overflow considered harmful? the impact of copy\&paste on android application security.
\newblock In {\em 2017 IEEE Symposium on Security and Privacy (SP)}, pages 121--136. IEEE, 2017.

\bibitem{fouad2018lawful}
Y~Fouad, A~Lodder, J~Hurdey, et~al.
\newblock A lawful basis for online proctoring, 2018.

\bibitem{greene2018platform}
Daniel Greene and Katie Shilton.
\newblock Platform privacies: Governance, collaboration, and the different meanings of “privacy” in ios and android development.
\newblock {\em New Media \& Society}, 20(4):1640--1657, 2018.

\bibitem{10.1145/3313831.3376511}
Hana Habib, Sarah Pearman, Jiamin Wang, Yixin Zou, Alessandro Acquisti, Lorrie~Faith Cranor, Norman Sadeh, and Florian Schaub.
\newblock "it's a scavenger hunt": Usability of websites' opt-out and data deletion choices.
\newblock In {\em Proceedings of the 2020 CHI Conference on Human Factors in Computing Systems}, CHI '20, page 1–12, New York, NY, USA, 2020. Association for Computing Machinery.

\bibitem{hamalainen2023evaluating}
Perttu H{\"a}m{\"a}l{\"a}inen, Mikke Tavast, and Anton Kunnari.
\newblock Evaluating large language models in generating synthetic hci research data: a case study.
\newblock In {\em Proceedings of the 2023 CHI Conference on Human Factors in Computing Systems}, pages 1--19, 2023.

\bibitem{horstmann2024those}
Stefan~Albert Horstmann, Samuel Domiks, Marco Gutfleisch, Mindy Tran, Yasemin Acar, Veelasha Moonsamy, and Alena Naiakshina.
\newblock “those things are written by lawyers, and programmers are reading that.” mapping the communication gap between software developers and privacy experts.
\newblock {\em Proceedings on Privacy Enhancing Technologies}, 2024.

\bibitem{horstmann2025sorry}
Stefan~Albert Horstmann, Sandy Hong, David Klein, Raphael Serafini, Martin Degeling, Martin Johns, Veelasha Moonsamy, and Alena Naiakshina.
\newblock {“Sorry for Bugging you so much.” Exploring Developers' Behavior Towards Privacy-Compliant Implementation}.
\newblock In {\em 2025 IEEE Symposium on Security and Privacy (SP)}, pages 1215--1233, 2025.

\bibitem{ico-guidance-controllers}
{Information Commissioner's Office}.
\newblock {Data controllers and data processors: what the difference is and what the governance implications are}, 2018.
\newblock \url{https://ico.org.uk/for-organisations/guide-to-data-protection/guide-to-the-general-data-protection-regulation-gdpr/controllers-and-processors/}.

\bibitem{jasmontaite2018data}
Lina Jasmontaite, Irene Kamara, Gabriela Zanfir-Fortuna, and Stefano Leucci.
\newblock Data protection by design and by default: Framing guiding principles into legal obligations in the gdpr.
\newblock {\em Eur. Data Prot. L. Rev.}, 4:168, 2018.

\bibitem{kamara2018understanding}
Irene Kamara and Paul De~Hert.
\newblock Understanding the balancing act behind the legitimate interest of the controller ground: A pragmatic approach.
\newblock {\em Brussels Privacy Hub}, 4(12), 2018.

\bibitem{kollnig2021fait}
Konrad Kollnig, Pierre Dewitte, Max Van~Kleek, Ge~Wang, Daniel Omeiza, Helena Webb, and Nigel Shadbolt.
\newblock A fait accompli? an empirical study into the absence of consent to third-party tracking in android apps.
\newblock In {\em Seventeenth Symposium on Usable Privacy and Security (SOUPS 2021)}, pages 181--196, 2021.

\bibitem{kuner2021eu}
Christopher Kuner, Lee~A Bygrave, Christopher Docksey, Laura Drechsler, and Luca Tosoni.
\newblock The eu general data protection regulation: A commentary/update of selected articles.
\newblock {\em Update of Selected Articles (May 4, 2021)}, 2021.

\bibitem{kyi2023investigating}
Lin Kyi, Sushil Ammanaghatta~Shivakumar, Franziska Roesner, Cristiana Santos, Frederike Zufall, and Asia Biega.
\newblock Investigating deceptive design in gdpr's legitimate interest.
\newblock In {\em Proceedings of the 2023 CHI Conference on Human Factors in Computing Systems}, pages 1--15, 2023.

\bibitem{google_fonts}
{LG München I (Regional Court Munich I)}.
\newblock {Verletzung des Persönlichkeitsrechts durch Datenschutzverstoß (Violation of the right of personality due to data protection infringement)}.
\newblock Available at \url{https://www.gesetze-bayern.de/Content/Document/Y-300-Z-BECKRS-B-2022-N-612?hl=true}, 2022.

\bibitem{li2021developers}
Tianshi Li, Elizabeth Louie, Laura Dabbish, and Jason~I Hong.
\newblock How developers talk about personal data and what it means for user privacy: A case study of a developer forum on reddit.
\newblock {\em Proceedings of the ACM on Human-Computer Interaction}, 4(CSCW3):1--28, 2021.

\bibitem{Liepia2019Claudette}
Rūta Liepiņa, Giuseppe Contissa, Kasper Drazewski, Francesca Lagioia, Marco Lippi, Hans~Wolfgang Micklitz, Przemyslaw Palka, Giovanni Sartor, and Paolo Torroni.
\newblock Gdpr privacy policies in claudette: Challenges of omission, context and multilingualism.
\newblock In {\em ASAIL@ICAIL}, 2019.

\bibitem{markham2012fabrication}
Annette Markham.
\newblock Fabrication as ethical practice: Qualitative inquiry in ambiguous internet contexts.
\newblock {\em Information, Communication \& Society}, 15(3):334--353, 2012.

\bibitem{matte2020purposes}
C{\'e}lestin Matte, Cristiana Santos, and Nataliia Bielova.
\newblock Purposes in iab europe’s tcf: which legal basis and how are they used by advertisers?
\newblock In {\em Annual Privacy Forum}, pages 163--185. Springer, 2020.

\bibitem{mchugh2012interrater}
Mary~L McHugh.
\newblock Interrater reliability: the kappa statistic.
\newblock {\em Biochemia medica}, 22(3):276--282, 2012.

\bibitem{NorwegianDPAdecision-MetaLI2023}
Norwwgian dpa against meta platfroms case 21/03530-16, 2023.

\bibitem{GDPRhub}
{noyb.eu}.
\newblock {GDPRhub}.
\newblock \url{https://gdprhub.eu/}, 2025.

\bibitem{C210/16}
European~Court of~Justice.
\newblock Case c‑210/16 wirtschaftsakademie schleswig-holstein, ecli:eu:c:2018:388, 2018.

\bibitem{CaseC40/17}
European~Court of~Justice.
\newblock Case c-40/17 fashion id gmbh \& co.kg v verbraucherzentrale nrw ev, ecli:eu:c:2019:629, 2019.

\bibitem{Planet49}
European~Court of~Justice.
\newblock {Case} c-673/17 verbraucherzentrale bundesverband v. planet49, ecli:eu:c:2019:801, 2019.

\bibitem{ICO-Guid-19}
Information~Commissioner's Office.
\newblock Guidance on the use of cookies and similar technologies, 2019.
\newblock \url{https://ico.org.uk/media/for-organisations/guide-to-pecr/guidance-on-the-use-of-cookies-and-similar-technologies-1-0.pdf}.

\bibitem{parnin2012crowd}
Chris Parnin, Christoph Treude, Lars Grammel, and Margaret-Anne Storey.
\newblock Crowd documentation: Exploring the coverage and the dynamics of api discussions on stack overflow.
\newblock {\em Georgia Institute of Technology, Tech. Rep}, 11, 2012.

\bibitem{parsons2023understanding}
Jonathan Parsons, Michael Schrider, Oyebanjo Ogunlela, and Sepideh Ghanavati.
\newblock {Understanding Developers Privacy Concerns Through Reddit Thread Analysis}.
\newblock {\em arXiv preprint arXiv:2304.07650}, 2023.

\bibitem{opinionads}
Article 29~Working Party.
\newblock Opinion 2/2010 on online behavioural advertising, 2010.

\bibitem{29WP-4-12-CookieExemption}
Article 29~Working Party.
\newblock Opinion 04/2012 on cookie consent exemption ({WP} 194), 2012.

\bibitem{judgment}
Article 29~Working Party.
\newblock Google spain sl and google inc. v agencia española de protección de datos (aepd) and mario costeja gonzález, 2014.

\bibitem{lawse}
Article 29~Working Party.
\newblock Guidelines on transparency under regulation 2016/679, wp260 rev.01, 2016.

\bibitem{proferes2021studying}
Nicholas Proferes, Naiyan Jones, Sarah Gilbert, Casey Fiesler, and Michael Zimmer.
\newblock Studying reddit: A systematic overview of disciplines, approaches, methods, and ethics.
\newblock {\em Social Media + Society}, 7(2):20563051211019004, 2021.

\bibitem{sanh2021multitask}
Victor Sanh, Albert Webson, Colin Raffel, Stephen~H Bach, Lintang Sutawika, Zaid Alyafeai, Antoine Chaffin, Arnaud Stiegler, Teven~Le Scao, Arun Raja, et~al.
\newblock Multitask prompted training enables zero-shot task generalization.
\newblock {\em arXiv preprint arXiv:2110.08207}, 2021.

\bibitem{Sant-etal-20-TechReg}
Cristiana Santos, Nataliia Bielova, and C{é}lestin Matte.
\newblock Are cookie banners indeed compliant with the law? deciphering {EU} legal requirements on consent and technical means to verify compliance of cookie banners.
\newblock {\em Technology and Regulation}, pages 91--135, 2020.

\bibitem{santos2021consent}
Cristiana Santos, Midas Nouwens, Michael Toth, Nataliia Bielova, and Vincent Roca.
\newblock Consent management platforms under the gdpr: processors and/or controllers?
\newblock In {\em Annual Privacy Forum}, pages 47--69. Springer, 2021.

\bibitem{senarath2018developers}
Awanthika Senarath and Nalin~AG Arachchilage.
\newblock Why developers cannot embed privacy into software systems? an empirical investigation.
\newblock In {\em Proceedings of the 22nd International Conference on Evaluation and Assessment in Software Engineering 2018}, pages 211--216, 2018.

\bibitem{shanmugam2022learning}
Divya Shanmugam, Fernando Diaz, Samira Shabanian, Mich{\`e}le Finck, and Asia Biega.
\newblock Learning to limit data collection via scaling laws: A computational interpretation for the legal principle of data minimization.
\newblock In {\em 2022 ACM Conference on Fairness, Accountability, and Transparency}, pages 839--849, 2022.

\bibitem{spiekermann2012challenges}
Sarah Spiekermann.
\newblock The challenges of privacy by design.
\newblock {\em Communications of the ACM}, 55(7):38--40, 2012.

\bibitem{stover2023website}
Alina St{\"o}ver, Nina Gerber, Henning Prid{\"o}hl, Max Maass, Sebastian Bretthauer, Matthias Hollick, Dominik Herrmann, et~al.
\newblock How website owners face privacy issues: Thematic analysis of responses from a covert notification study reveals diverse circumstances and challenges.
\newblock {\em Proceedings on Privacy Enhancing Technologies}, 2023.

\bibitem{tahaei2022privacy}
Mohammad Tahaei, Julia Bernd, and Awais Rashid.
\newblock Privacy, permissions, and the health app ecosystem: A stack overflow exploration.
\newblock In {\em Proceedings of the 2022 European Symposium on Usable Security}, pages 117--130, 2022.

\bibitem{tahaei2021privacy}
Mohammad Tahaei, Alisa Frik, and Kami Vaniea.
\newblock Privacy champions in software teams: Understanding their motivations, strategies, and challenges.
\newblock In {\em Proceedings of the 2021 CHI Conference on Human Factors in Computing Systems}, pages 1--15, 2021.

\bibitem{tahaei2022understanding}
Mohammad Tahaei, Tianshi Li, and Kami Vaniea.
\newblock Understanding privacy-related advice on stack overflow.
\newblock {\em Proc. Priv. Enhancing Technol.}, 2022(2):114--131, 2022.

\bibitem{tahaei2020understanding}
Mohammad Tahaei, Kami Vaniea, and Naomi Saphra.
\newblock Understanding privacy-related questions on stack overflow.
\newblock In {\em Proceedings of the 2020 CHI Conference on Human Factors in Computing Systems}, pages 1--14, 2020.

\bibitem{Taylor-Hiscock_2021_GDPR_Compliance}
Robb Taylor-Hiscock.
\newblock {Your complete guide to General Data Protection Regulation (GDPR) compliance}.
\newblock {\em {OneTrust Blog}}, April 2021.

\bibitem{ePD-09}
European Union.
\newblock Directive 2009/136/ec of the european parliament and of the council, 2009.

\bibitem{vasilescu2013stackoverflow}
Bogdan Vasilescu, Vladimir Filkov, and Alexander Serebrenik.
\newblock Stack overflow and github: Associations between software development and crowdsourced knowledge.
\newblock In {\em 2013 International Conference on Social Computing}, pages 188--195. IEEE, 2013.

\bibitem{veale2018algorithms}
Michael Veale, Reuben Binns, and Lilian Edwards.
\newblock Algorithms that remember: model inversion attacks and data protection law.
\newblock {\em Philosophical Transactions of the Royal Society A: Mathematical, Physical and Engineering Sciences}, 376(2133):20180083, 2018.

\bibitem{Webley-qualit-content-analysis}
Lisa Webley.
\newblock {\em Qualitative approaches to empirical legal research}, pages 926--948.
\newblock Oxford Handbooks. Oxford University Press, United Kingdom, November 2010.

\bibitem{stackoverflowjobs}
Tom Winter.
\newblock How to source developers from stack overflow, 2021.

\bibitem{wu2019developers}
Yuhao Wu, Shaowei Wang, Cor-Paul Bezemer, and Katsuro Inoue.
\newblock How do developers utilize source code from stack overflow?
\newblock {\em Empirical Software Engineering}, 24(2):637--673, 2019.

\end{thebibliography}
\end{document}